\documentclass{aa}

\usepackage{graphicx}
\usepackage{txfonts}
\usepackage[dvipsnames]{xcolor}
\usepackage{xparse,xcolor}

\begin{document}

   \title{The disappearance and reformation of the accretion disc during a low state of FO Aquarii}

   \author{J.-M. Hameury
          \inst{1,4}
          \and
          J.-P. Lasota\inst{2,3,4}
          }

   \institute{Université de Strasbourg, CNRS, Observatoire Astronomique de Strasbourg, UMR 7550, 67000 Strasbourg, France\\
             \email{jean-marie.hameury@astro.unistra.fr}
        \and
        Institut d'Astrophysique de Paris, CNRS et Sorbonne Universit\'es, UPMC Paris~06, UMR 7095, 98bis Bd Arago, 75014 Paris, France
		\and
        Nicolaus Copernicus Astronomical Center, Polish Academy of Sciences, Bartycka 18, 00-716 Warsaw, Poland
		\and               
		Kavli Institute for Theoretical Physics, Kohn Hall, University of California, Santa Barbara, CA 93106, USA
}

   \date{Received / Accepted}

  \abstract
  {FO Aquarii, an asynchronous magnetic cataclysmic variable (intermediate polar) went into a low-state in 2016, from which it slowly and steadily recovered without showing dwarf nova outbursts. This requires explanation since in a low-state, the mass-transfer rate is in principle too low for the disc to be fully ionized and the disc should be subject to the standard thermal and viscous instability observed in dwarf novae.}
   {We investigate the conditions under which an accretion disc in an intermediate polar could exhibit a luminosity drop of 2 magnitudes in the optical band without showing outbursts.}
   {We use our numerical code for the time evolution of accretion discs, including other light sources from the system (primary, secondary, hot spot).}
   {We show that although  it is marginally possible for the accretion disc in the low-state to stay on the hot stable branch, the required mass-transfer rate in the normal state would then have to be extremely high, of the order of 10$^{19}$ g~s$^{-1}$ or even larger. This  would make the system so intrinsically bright that its distance should be much larger than allowed by all estimates. We show that observations of FO Aqr are well accounted for by the same mechanism that we have suggested as explaining the absence of outbursts during low states of VY Scl stars: during the decay, the magnetospheric radius exceeds the circularization radius, so that the disc disappears before it enters the instability strip for dwarf nova outbursts. }
   {Our results are unaffected, and even reinforced, if accretion proceeds both via the accretion disc and directly via the stream during some intermediate stages; the detailed process through which the disc disappears still needs investigations.}


   \maketitle
%

\section{Introduction}

Intermediate polars (IPs)  are cataclysmic variables (CVs), in which a magnetic white dwarf accretes matter from a Roche-lobe filling low-mass companion; the magnetic field of the white dwarf is strong enough to disrupt the disc and to channel the accretion flow onto the magnetic poles. As a result the luminosity is modulated at the white dwarf's spin period. The strength of magnetic field, however, is not sufficient to synchronize the white dwarf spin with the orbital rotation as in AM Her stars (see \citealt{w95} for a comprehensive description  of cataclysmic variables, and \citealt{w14} for a review of magnetic CVs).

The existence of an accretion disc in IPs has been controversial. \citet{hkl86} argued that in some systems, the white dwarf magnetic field is large enough to prevent the formation of an accretion disc, and that the accretion stream originating from the $L_1$ point would then impact directly onto the white dwarf's magnetosphere. There is observational evidence that many IPs do posses accretion discs \citep{h91}. In \object{V2400 Oph} \citep{hb02}, however, the emission is pulsed at the beat frequency $\omega-\Omega$ between the orbital frequency $\Omega$ and the spin frequency $\omega$, indicating that accretion flips from one pole to the other as expected when no disc is present and the stream directly interacts with the magnetosphere. Other systems, including FO Aqr \citep{h93}, show pulsations both at the spin frequency and at the beat frequency. These systems have been interpreted as a mix of disc and direct accretion resulting from the accretion stream overflowing the disc.

The existence of an accretion disc is also revealed by the presence of dwarf nova outbursts, which are due to a thermal-viscous instability occurring when, somewhere in the disc, the temperature is close to the hydrogen ionization temperature and opacities become a steep function of temperature \citep[see][for a review of the disc instability model]{l01}. In the disc instability model (DIM) describing dwarf-nova outbursts,  the plot of the relation between the temperature at a given point in the disc, or equivalently the local mass-transfer rate, and the corresponding local surface density $\Sigma$ has an S-shape; the upper hot, and the lower, cool branches are stable, while the intermediate segment of the S-curve is thermally and viscously unstable. This intermediate branch corresponds to partial ionization of hydrogen, and high sensitivity of the opacities to variations of the temperature. Only systems accreting at a rate high enough that hydrogen is everywhere ionised in the disc, or low enough that hydrogen is entirely neutral, can be steady. Whereas for non-magnetic CVs the latter option requires very low, unrealistic accretion rates, it has been shown by \citet{hl17b} that it can explain why so few IPs exhibit dwarf-nova outbursts. 

\object{FO Aqr} is an IP that underwent a low-state in 2016 (decrease of the optical luminosity by 2 magnitudes, from a normal state with $V = 13.5$ to a faint state with $V=15.5$), from which it recovered slowly and steadily over a time scale of several months \citep{lgk16}. X-ray observations \citep{kgl17} show that the mass-transfer rate decreased by about one order of magnitude. Due to solar conjunction, not much data is available for the descent to the low-state, which has been rapid (time scale of weeks) as compared to the recovery time. Interestingly, no dwarf nova outburst has been detected during the rise from the low-state which analogous of the behaviour of VY Scl stars in a similar phase of their luminosity variations. VY Scl stars are bright cataclysmic variables which occasionally enter low-states fainter by one magnitude or more; although such drops in luminosity should bring these systems in the instability strip, they show no outbursts. This is puzzling, because one would expect that in the low-state, the mass-transfer is low enough for the disc to become thermally and viscously unstable. \citet{lhk99} suggested that irradiation of the accretion disc by a hot white dwarf prevents the occurrence of dwarf nova outbursts. However, \citet{hl02} showed that irradiation cannot completely suppress the outbursts (especially during the long rise from minimum) and suggested instead that VY Scl stars are magnetic CVs, in which the magnetic field is strong enough to lead to disc disappearance during the low state. The required magnetic moments are comparable to that observed in IPs. It is therefore tempting to apply the same idea to confirmed IPs when they show no-outburst behaviour analogous to that of VY Scl stars.

In this paper, we consider the specific case of FO Aqr. This source possesses an accretion disc during the high state, as shown by its partial eclipses. We examine in turn the three possibilities that could account for the absence of dwarf nova outbursts in FO Aqr: a very high mass-transfer rate, for which the disc can stay on the hot branch during the low-state, a very low mass-transfer for which the disc sits on the cool branch even during the high state, and finally the case of a mass-transfer rate corresponding to the system's orbital period and show that in this case the accretion disc, disrupted by the white dwarf's magnetic field, vanishes in the low-state before entering the instability strip.

We adopt the following parameters for FO Aqr:
\begin{itemize}
\item Orbital period: 4.85 h; spin period: 1254s \citep{ps83,kgb16}.
\item Orbital inclination $i$: FO Aqr has a grazing eclipse; we take $i=70\degr$ \citep{hmc89}.
\item Primary mass $M_1$: this parameter is poorly constrained. Values found in the literature range from 0.45 M$_\odot$\citep{ynm10} to 1.22 M$_\odot$ \citep{crw98}. Here, we adopt $M_1=0.7$ M$_\odot$, close to the value of 0.61 M$_\odot$ quoted by \citet{bga09} from the Swift/BAT survey.
\item Secondary mass $M_2$: we take $M_2$ = 0.4 M$_\odot$, appropriate for a system with a period of 4.85 h in which the secondary is not evolved. This value is slightly smaller than the 0.46 M$_\odot$ deduced from the empirical mass-period relation derived by \citet{k06}; deviations and scatter from this mass-period relation are expected as a result of differences in primary masses and in evolutionary tracks.
\item Secondary effective temperature: \citet{dbmm94} found that the secondary effective temperature is in the range 3700 -- 3900 K; here we adopt a value of 3800 K for the unirradiated part of the secondary. This corresponds to an M0 star, typical of what is expected for a period of 4.85 h.
\item Distance $d$: this parameter is also poorly constrained. \citet{pm14} assume $d=450^{+240}_{-160}$ pc, which corresponds to an uncertainty by a factor 2-3 in the luminosity. 
\item Mass transfer rate: this can be estimated either by converting the X-ray luminosity into an accretion rate, or by using spectral models. The second method is model dependent, while the first one suffers from the uncertainty on the source distance, and from the uncertainty on the bolometric correction. There is indeed a large intrinsic absorption in IPs; the total accretion luminosity is typically 50 -- 100 times larger than the 2-10 keV luminosity \citep[we refer to][for a discussion of this effect]{w95}. The 2-10 keV X-ray flux given in  Koji Mukai's on-line catalogue\footnote{https://asd.gsfc.nasa.gov/Koji.Mukai/iphome/iphome.html} is $3.5 \times 10^{-11}$ erg~cm$^{-1}$~s$^{-1}$. For the assumed primary mass and distance, this corresponds to accretion rates of the order of $4 - 8 \times 10^{17}$ g~s$^{-1}$, with large uncertainties. \citet{mio94} quote a small value of the bolometric luminosity ($4.6 \times 10^{33}$ erg~s$^{-1}$ for a distance of 400 pc). This would imply very low mass transfer rates ($4 \times 10^{16}$ g~s$^{-1}$), too low to account for the properties of FO Aqr (see below). On the other hand, \citet{dbmm94} estimate a mass accretion rate of $9.2 \times 10^{17}$ g~s$^{-1}$ when fitting the fraction of the optical and infrared flux that is not modulated with the orbital period with a disc plus blackbody component, implying a larger absorption of the X-ray flux than estimated by \citet{mio94}. In the low state, one expects the mass transfer rate to be reduced by about a factor 10 from the high state.
\item Magnetic moment: \citet{bkn09} found some evidence for circular polarization in the I band, and variation of this with spin phase. This detection is marginal, however, and assuming a surface magnetic field of a few MG as for other intermediate polars for which circular polarization has been detected, one would get a magnetic moment of a few $10^{32}$ -- 10$^{33}$ G~cm$^3$ for a 0.7 M$_\odot$ white dwarf. \citet{nws04} suggested a magnetic moment of $1.2 \times 10^{33}$ G cm$^3$ for FO Aqr, under the assumption of spin equilibrium, when the rate at which angular momentum is accreted by the white dwarf is balanced by the braking effect of the magnetic torque on the disc at its inner edge.
\end{itemize}

\section{The case for a low accretion rate system}

   \begin{figure}
   \centering
   \includegraphics[angle=-90,width=\columnwidth]{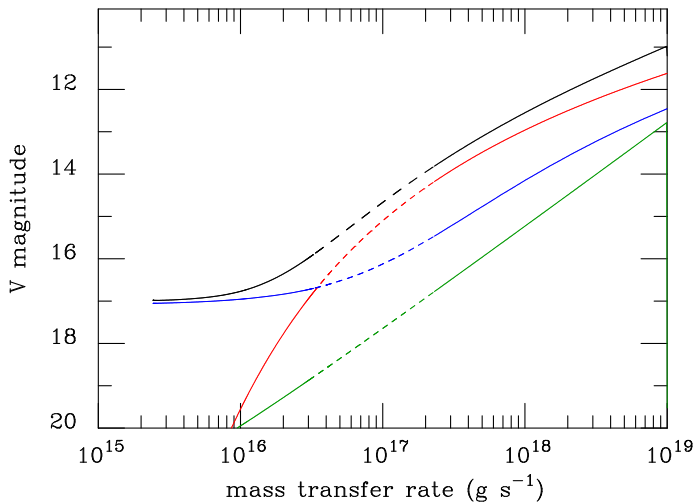}
   \caption{Visual magnitude for a system with the parameters of FO Aqr (see text), as a function of the mass-transfer rate from the secondary. The thick solid curve is the total luminosity; the thin red curve is the contribution from the {\rm disc}, the blue one that of the secondary star, and the green one that of the hot spot. The white dwarf at a temperature of 10$^4$ K corresponds to a magnitude $m_V=20.7$. The dashed portion of these curves correspond to unstable cases, in which a fraction of the disc sits on the intermediate branch of the S curve and is therefore not physical.}
   \label{fig:vmag1}
   \end{figure}
   
   \begin{figure}
   \centering
   \includegraphics[angle=-90,width=\columnwidth]{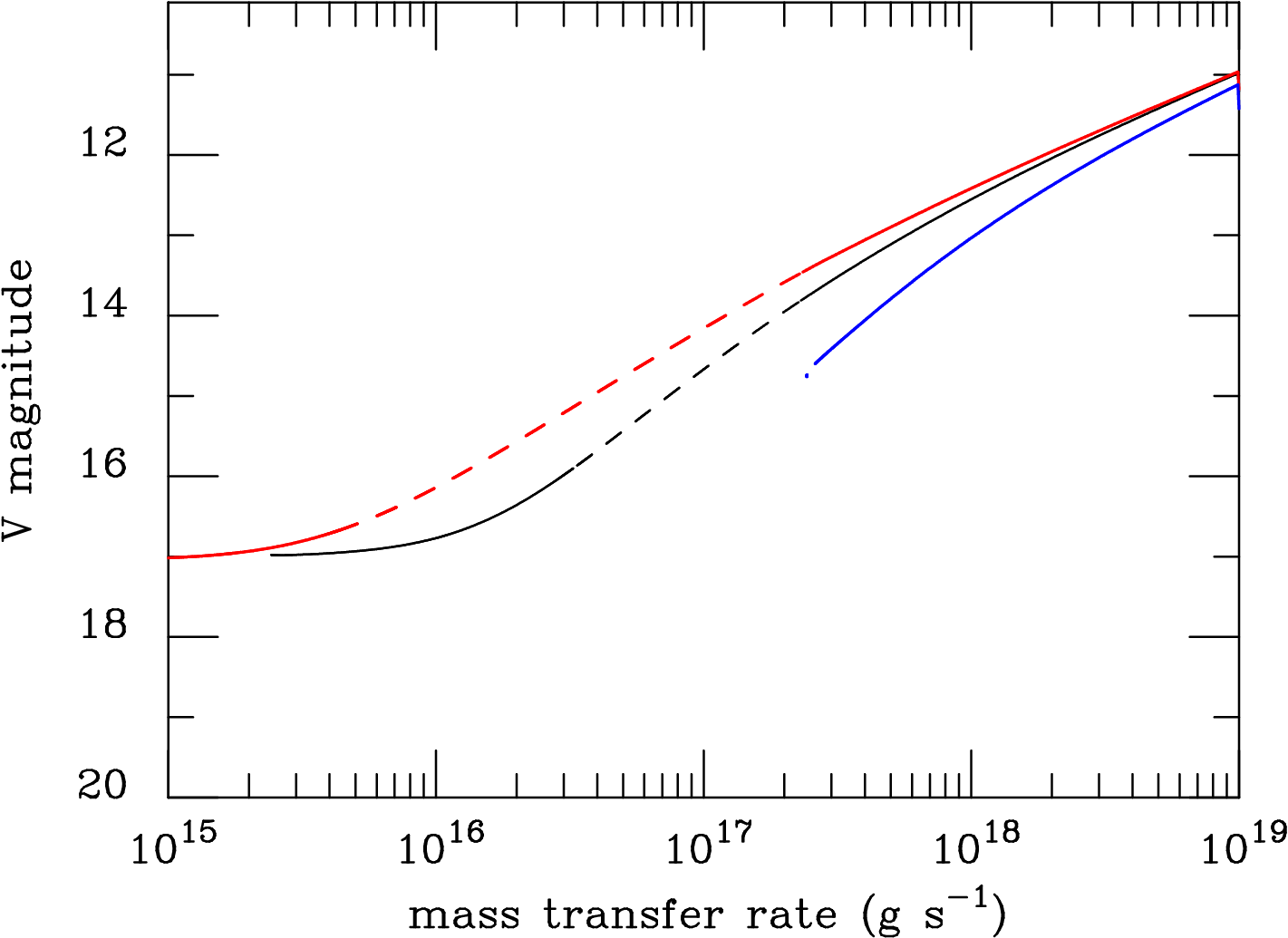}
   \caption{Visual magnitude for a system with the parameters of FO Aqr (see text), as a function of the mass-transfer rate from the secondary, for $\mu_{30}$ = 10 (red curve), 100 (black curve), 1000 (blue curve). The dashed portion of these curves correspond to unstable cases.}
   \label{fig:vmag2}
   \end{figure}
   
   \begin{figure}
   \centering
   \includegraphics[angle=-90,width=\columnwidth]{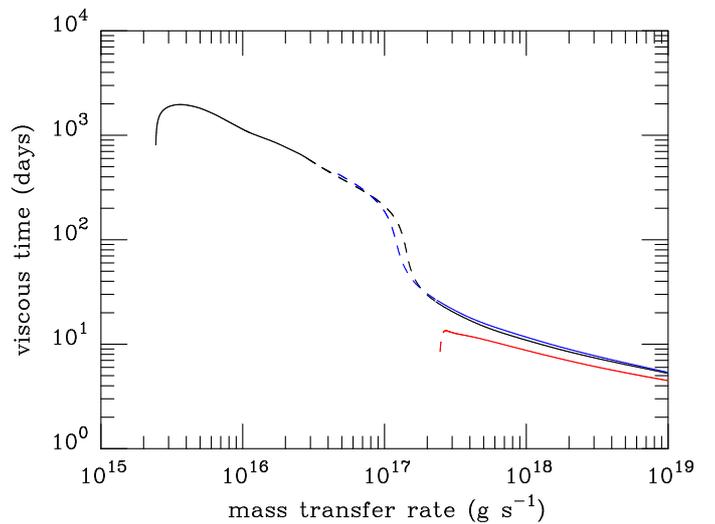}
   \caption{Viscous time for a system with the parameters of FO Aqr (see text), as a function of the mass-transfer rate from the secondary, for $\mu_{30}$ = 10 (red curve), 100 (black curve), 1000 (blue curve). The dashed portion of these curves correspond to unstable cases.}
   \label{fig:viscous_time}
   \end{figure}

A system will not undergo dwarf nova outbursts during the decay to a low-state if, in the normal, high state, the disc already sits on the cold, stable branch. For this to happen, the mass-accretion rate must be everywhere in the disc lower than the maximum value $\dot{M}_A$ allowed for a cold, stable disc.  $\dot{M}_A$ is an increasing function of radius, and reaches its minimum at the inner edge of the disc, given by the magnetospheric radius $r_{\rm mag}$
\begin{equation}
r_{\rm mag} = 2.66 \times 10^{10} \mu_{33}^{4/7} M_1^{-1/7} \left( \frac{\dot{M}_{\rm acc}}{10^{16} \; \rm g s^{-1}} \right)^{-2/7} \; \rm cm,
\label{eq:rmag}
\end{equation}
where $\dot{M}_{\rm acc}$ is the accretion rate onto the white dwarf, and $\mu_{33}$ is the white dwarf magnetic moment in units of $10^{33}$ G\,cm$^3$; the white dwarf mass $M_1$ is measured in solar units. This is possible if the mass-transfer from the secondary is less than \citep{hl17b}
\begin{equation}
\dot{M}_{\rm tr} < 4.89 \times 10^{16} M_1^{-0.72} \mu_{33}^{0.87} \; \rm g s^{-1}.
\label{eq:mdotmax}
\end{equation}

Equation (\ref{eq:mdotmax}) must be satisfied both during the normal and low-state. We have used our DIM evolution code as described in \citet{hl17a} and \citet{hl17b} to estimate the visual magnitude of a steady disc. As we are interested in steady state solutions, we take arbitrary large values for the time step. We include the contributions to the optical light from the primary star, for which we take a temperature $T_1 = 10^4$ K. The secondary star is treated as described in \citet{shl03}; the secondary is divided in two hemispheres with different effective temperatures, an non-irradiated one with $T_{\rm eff, unirr} = 3800$ K, and an irradiated one for which 
\begin{equation}
\label{eq:t2}
\sigma T_{\rm eff,irr}^4 = \sigma T_{\rm eff, unirr}^4  + (1-\eta) F_{\rm irr}
\end{equation}
where $\eta$ is the albedo, taken to be 0.8, and $F_{\rm irr}$ is the irradiation flux due to accretion onto the white dwarf, estimated at distance $a$, $a$ being the orbital separation. Specifically, one has:
\begin{equation}
\label{eq:firr}
4 \pi a^2 F_{\rm irr} =  \frac{G M_1 \dot{M}_{\rm acc}}{R_1} + 4 \pi R_1^2 T_1^4
\end{equation}
where $R_1$ is the white dwarf radius. $T_{\rm eff, unirr}$ corresponds to an M0 star, typical of what is expected for a period of 4.85 h. For an accretion rate of $10^{18}$ g~s$^{-1}$, the effective temperature of the irradiated secondary is 7560 K; for an accretion rate of $6 \times 10^{17}$ g~s$^{-1}$, this temperature is 6700 K. These compare well with the values obtained by \citet{dbmm94}; these authors fitted the spectrum of the optical and infrared light modulated at the orbital period and found that it can be fitted by a cool component, presumably representing the irradiated emission with a temperature of 6800 K, and by a hot component with a temperature $ \geq 20000$ K, with a large uncertainty. This hot component is presumably due to the hot spot. Using HST and IUE observations, \citet{dsb99} were able to constrain this temperature to a more accurate value of $19500 \pm 500$ K. The contribution of the hot spot is also included in our model; we assume a colour temperature of 10$^4$ K, typical of hot spots in CVs, but smaller than the value found by \citet{dsb99}. This possibly overestimates the contribution from the hot spot in the optical, but as will be seen below, the hot spot contribution to the optical light is small, so that this should not introduce any significant error.

Figure \ref{fig:vmag1} shows the magnitude of a system with the parameters of FO Aqr, as a function of the mass-transfer from the secondary, for a magnetic moment of 10$^{32}$ G~cm$^3$. The dashed portions of the curves correspond to unstable discs, and are therefore not physical; the curves are also interrupted at low mass-transfer rates, when the magnetospheric radius becomes equal to the circularization radius and the disc can no longer exist. The circularization radius is the radius of a Keplerian orbit with the same specific angular momentum as the transferred matter had when leaving $L_1$ \citep{fkr02}; for the parameters adopted here for FO Aqr, this is $0.107 a \sim 1.12 \times 10^{10}$ cm, $a$ being the orbital separation. As can be seen, the system luminosity is dominated by the irradiated secondary at low mass-transfer rates, and the difference between the maximum possible brightness of a system with a disc on the cool branch and the minimum corresponding to very small accretion rates is only 1 mag for the orbital parameters of FO Aqr and for $\mu=10^{32}$ G~cm$^3$. This conclusion is independent of  the assumed source distance; the maximum amplitude depends only on the secondary luminosity and on the magnetic moment of the white dwarf. Figure \ref{fig:vmag2} shows the same as Fig. \ref{fig:vmag1} for different values of $\mu$; as can be seen, the maximum amplitude increases with increasing $\mu$, but is limited by the fact that the magnetospheric radius cannot exceed the circularization radius. Variations as large as 2 mag cannot therefore be accounted for if the disc is to stay on the cool stable branch. In addition, the visual magnitudes for a system on the cold branch are faint: the maximum brightness is 15.5 -- 16, depending on $\mu$, which is 2 mag fainter than FO Aqr in the normal state. In order to reconcile the predicted and observed values, one would have to  reduce the distance by a factor 2.5 at least, which, despite the large uncertainty on the distance, is unlikely. In addition, at those low mass transfer rates, the effective temperature of the illuminated side of the secondary is small; for $\dot{M}_{\rm acc} = 4 \times 10^{16}$g~s$^{-1}$, $T_{\rm eff,irr}=4270$ K, much smaller than the value found by \citet{dbmm94}

It is also worth noting that, for $\mu=10^{32}$ G~cm$^3$, we find from Fig. \ref{fig:vmag2} that the maximum mass-transfer rate for a system on the cold stable branch is $2.8 \times 10^{16}$ g~s$^{-1}$, which is larger than the value given by Eq. \ref{eq:mdotmax} by a factor 3. This difference arises from the fact that the inner disc radius is not very different from the outer disc radius (a factor $\sim 5$); the boundary-condition term $ [1 - (r_{\rm mag}/r)^{1/2}]$ which enters in the steady state solution is important and introduces significant  corrections which are not well captured by the fits in \citet{hl17b}.

We have assumed that the disc is steady, which is valid only if the viscous time
\begin{equation}
t_{\rm visc} = \max (2 \pi r^2 \frac{\Sigma}{\dot{M}})
\end{equation}
is less than the time-scale on which the mass-transfer rate from the secondary evolves. Figure \ref{fig:viscous_time} shows $t_{\rm visc}$ as a function of the mass-transfer rate for a steady disc, for various values of the white dwarf magnetic moment. As can be seen, $t_{\rm visc}$ is of the order of   500 - 1000 days for $\mu=10^{32}$ G~cm$^3$ on the cold branch of the S curve, longer than the duration of the low-state in FO Aqr. We therefore expect the disc not be steady. The disc will be slightly more luminous in the low-state than a steady disc with the same mass-transfer rate, thereby reducing the amplitude of the luminosity variation. One should also note that the maximum of the viscous time is reached in the outer parts of the accretion disc, explaining why $t_{\rm visc}$ depends only weakly on $\mu$. If $\mu$ becomes large, the effect of the inner boundary condition is felt  throughout the entire disc, which explains why the curve calculated for $\mu = 10^{33}$ G~cm$^3$ deviates from the other two. 

   \begin{figure}
   \centering
   \includegraphics[angle=-90,width=\columnwidth]{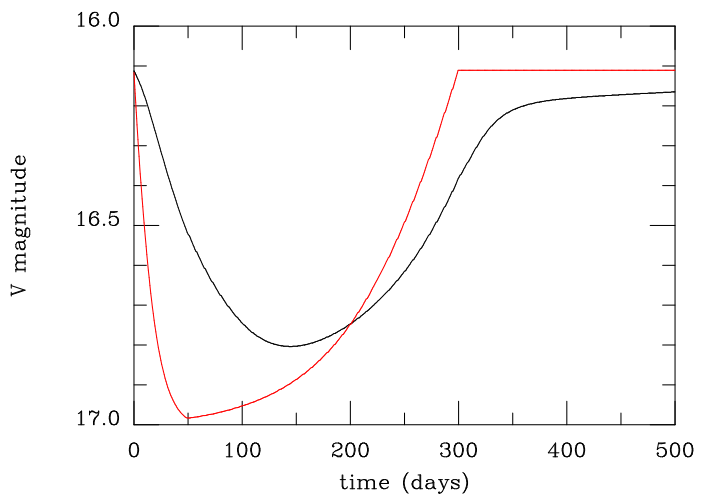}
   \caption{Light curve of a system staying on the cold branch, in which the mass-transfer rate decreases exponentially by a factor 10 during the first 50 days, then recovers on a time scale of 250 days. The black curve is the actual V magnitude, the red one correspond to a steady state system.}
   \label{fig:evol}
   \end{figure}
   
Figure \ref{fig:evol} shows the evolution of a system with an initial mass-transfer rate of $2.64 \times 10^{16}$ g~s$^{-1}$ that decreases exponentially by a factor 10 during 50 days, and then recovers on a 250 days time scale. As can be seen, the departure from steady state is significant, and the amplitude of the optical luminosity variations is smaller than expected from the steady state assumption. It is also difficult to account for the rapid decline observed in FO Aqr.

We therefore conclude that the normal and low-states of FO Aqr cannot be accounted for by a disc residing permanently on the cold branch of the S-curve.

\section{The case for a high accretion rate system}

We now examine the possibility that the disc can be on the hot stable branch both during the normal and the low states. This is possible if the mass-transfer rate from the secondary is always larger than the critical value for the disc to stay on the hot branch at the outer edge \citep{hl17b}
\begin{equation}
\dot{M}_{\rm tr} > \dot{M}_B (0.8 r_{\rm out}) = 9.5 \times 10^{15} \left( \frac{r_{\rm out}}{10^{10} \rm cm} \right)^{2.65} \; \rm g s^{-1},
\label{eq:mdotmin}
\end{equation}
where we have neglected the small dependence on the viscosity parameter $\alpha$. The $0.8 R_{\rm out}$ term accounts for additional  heating terms in the outer disc (heating due to the dissipation of tidal torques and by the stream impact on the disc edge, \citealt{bhl01}), so that the disc is stable for mass-transfer rates smaller than $\dot{M}_B(r_{\rm out})$. For FO Aqr, the outer disc radius is $3.7 \times 10^{10}$ cm, and the critical rate is $3.2 \times 10^{17}$ g~s$^{-1}$. Note that this rate depends only on the primary mass and is therefore accurate to better than a factor 2. 

The examination of Fig. \ref{fig:vmag2} again shows the maximum amplitude that one can expect for a disc on the hot branch. The mass-transfer for which the disc becomes unstable depends slightly on the magnetic moment, whereas Eq. (\ref{eq:mdotmin}) does not depend on $\mu$; the reason for this is that the inner boundary condition$\nu\Sigma=0$, where $\nu$ is the kinematic viscosity and $\sigma$ the surface density introduces a term $1-(r/r_{\rm in})^{1/2}$, $r_{\rm in}$ being the inner disc radius, that slightly modifies the outer structure of the disc and becomes increasingly important as the magnetospheric truncation radius increases. Figure \ref{fig:vmag2} shows that 2 mag amplitudes are possible if the mass-transfer rate in the normal state is at least of the order of a few times $10^{18}$ or $10^{19}$ g~s$^{-1}$, and if it is of order of $5 \times 10^{17}$ g~s$^{-1}$ in the low-state. These lower limits are obtained for large values of $\mu$ and correspond to the case where the disc has almost disappeared in the low state, and therefore require some fine tuning. These rates are much higher than the rates expected from secular evolution models \citep{kbp11}, and would imply that the current mass-transfer deviate by 2 orders of magnitude from the secular mean; this is in principle possible, though. The system would also have to be brighter by 2 magnitudes than observed for the fiducial distance of 450 pc; the distance would then have to be larger than 1 kpc to reconcile the model and the observations. Given the galactic latitude of FO Aqr ($b=-49.16 \deg$), this would put the source at a distance larger than 750 pc above the galactic plane.

The examination of Fig. \ref{fig:viscous_time} shows that when the disc is on the hot, stable branch of the S-curve, viscous time is of the order of a few days, much shorter than the time scale on which the system luminosity has been observed to vary, and the steady state assumption is fully justified.

We therefore conclude that it is possible, but very unlikely, that the disc of FO Aqr was in the hot branch of the S curve during its low-state.

\section{The case for the disappearance of the accretion disc}

   \begin{figure}
   \centering
   \includegraphics[angle=-90,width=\columnwidth]{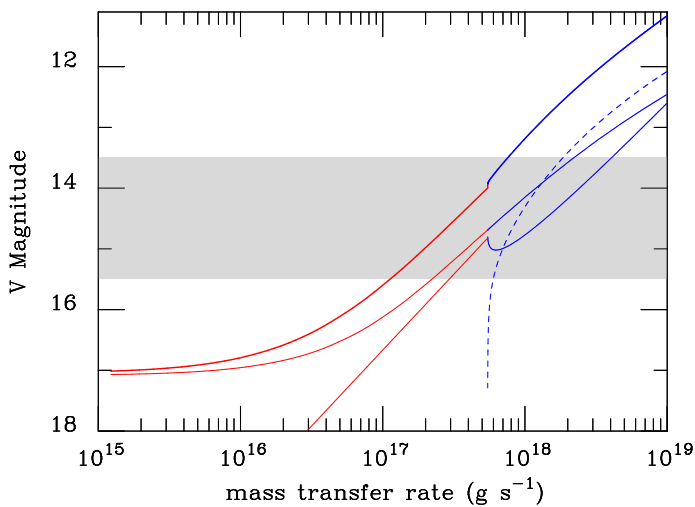}
   \caption{Optical magnitude for a system with the parameters of FO Aqr (see text) and $\mu = 1.5 \times 10^{33}$ G~cm$^3$, as a function of the mass-transfer rate. The thick solid line is the total optical luminosity; the thin dashed curve represents to the disc contribution and the upper and lower thin solid curves represent to the secondary and hot spot contributions respectively. The red portion of the curve corresponds to discless accretion. The grey area corresponds to the optical luminosity variations between the normal and low-states.}
   \label{fig:FO}
   \end{figure}

The last option is that the disc is on the hot branch of the S-curve during the normal state, and that, during decline, the magnetospheric radius exceeds the circularization radius before the critical $\dot{M}_{\rm B}$ is reached. The mass-transfer rate in the normal state should then be of the order of 10$^{18}$ g~s$^{-1}$ (the minimum value would be $5.5 \times 10^{17}$ g~s$^{-1}$, but this would imply that the disc has almost disappeared in the normal state) and the visual magnitude should be $\sim 13.5$, as observed, and the magnetic moment should be somewhat larger than 10$^{33}$ G~cm$^3$, so that the disc can vanish before entering the instability strip.

Because the disc, when it exists, sits on the hot stable branch, its viscous time is short, and the steady state assumption is fully justified. 

Figure \ref{fig:FO} shows the optical luminosity of a system in steady state with the parameters of FO Aqr, and a magnetic moment of $1.5 \times 10^{33}$ G~cm$^3$ as a function of the mass-transfer rate from the secondary. The disc, when it exists, is always in the hot state; when the mass-transfer rate decreases below $5.5 \times 10^{17}$ g~s$^{-1}$, the magnetospheric radius exceeds the circularization radius, and the disc vanishes. The optical luminosity is then due to the contributions of the irradiated secondary, of the primary, and of the region where the stream of matter leaving the $L_1$ point impacts the magnetosphere. For simplicity, we assume that the stream kinetic energy is dissipated entirely in this region and that the effective temperature is $10^4$ K;  when the disc vanishes, this contribution is thus equivalent to that of the hot spot, and then decreases as the mass transfer rate decreases. The hot spot contribution increases slightly just before the disappearance of the disc; this is due to the fact that the outer disc radius decreases until reaching the circularization radius, at which point the disc disappears. The heated secondary and the stream impact onto the magnetosphere dominate the optical light and have similar contributions except for very low mass transfer rates.

This scenario therefore provides a very reasonable explanation of the low-state behaviour of FO Aqr. 

\section{Conclusions}

We have shown that the absence of dwarf nova outbursts during the decline and slow recovery of FO Aqr is naturally accounted for if the magnetic moment of the white dwarf is large enough for the accretion disc to disappear before the mass-transfer rate enters the instability strip. It is, however, possible that the disc is always present and hot, but this would require mass-transfer rates much larger than expected, and would imply large luminosities which could be reconciled with the observed optical magnitude only if the distance to FO Aqr is larger by a factor at least 2 than normally assumed. Given the source position, this would put the source 750 pc above the galactic plane, which would be very difficult to account for.

The disappearance of the accretion disc has observable consequences. It is interesting to note that the partial eclipse shape changed during the low-state; the eclipses became narrower and less deep, indicating that the accretion disc has shrunk \citep{lgk16}. As the phased light curve over the orbital period was calculated by \citet{lgk16} over their entire dataset, it is not possible to decide from these observations if the accretion disc completely disappeared at some point, but one nevertheless expects that material impacting the magnetosphere should form some structure in the orbital plane, possibly mimicking a ring whose size should be comparable to the magnetospheric radius, which is smaller by a factor $\sim$ 2 than the \rm{disc} size in the normal state; this is about the ratio of the eclipse FWHM in the normal and low-states. 

One also expects that the power spectrum in the optical to change when the disc disappears. \citet{wk92} showed that discless accretion models produce a rich variety of power spectra, and that the presence of beat-frequency modulations is incompatible with dominant accretion via a disc. In the case of discless accretion, the power spectrum should be dominated by the beat frequency $\omega - \Omega$ between the spin and orbital frequency and its first harmonic $2\omega - 2\Omega$, whereas in the case of disc accretion, the spin frequency should be dominant \citep{fw99}. \citet{lgk16} found that, in the low-state, the beat component amplitude is more than twice that of the spin component, while in the hight state, it in only 20\% of the spin component. Although the modelling of the power spectrum is complex because of the complex geometry of the accretion flow and of the contribution of several sources to the optical light (see above), the change in the power spectrum clearly supports our finding that the disc disappears during the low-state.

There have been suggestions that accretion in IPs can occur both via an accretion disc and directly onto the white dwarf; this would happen if, for example, the stream leaving the $L_1$ point overflows the accretion disc after impacting its outer edge \citep{al96}. This was found to occur in FO Aqr at some epochs \citep{h93}. However, our conclusions that the disc must disappear during the faintest portion of the low-state would still hold. The absence of dwarf nova outburst implies that the disc, when it exists, must stay on the hot branch. Since only a fraction of the accretion flow is processed through the accretion disc, this makes it even more difficult to account for a an accretion \rm{disc} in the low-state: the system brightness in the high state would have to be even larger than estimated above. This does not precludes the possibility that the transition between disc and discless accretion could be gradual, and that both may happen at some time, but the amplitude of the observed variations makes it most likely that the disc has to vanish in the lowest state. The way the disc disappears has been treated in a very simplified manner in this paper. However, the study of the detailed precesses through which the disc disappears would require 3D simulations of the accretion flow which are clearly outside the scope of this paper.

We finally note that the reasons for  mass-transfer variations in cataclysmic variables are not understood at present. Low-states are observed in magnetic CVs, and most notably in AM Her systems which do not possess accretion discs, so that luminosity variations directly reflect variations in the mass-transfer rate from the secondary. \citet{lp94} proposed that variations in VY Scl stars could be due to star spots passing in the $L_1$ region, and \citet{hgm00} later extended and developed this model to AM Her systems; but as noted by \citet{lgk16}, the unusually long duration of the recovery in FO Aqr remains somewhat mysterious.

\begin{acknowledgements}
This work was supported by a National Science Centre, Poland grant 2015/19/B/ST9/01099 and by the National Science Foundation under Grant No. NSF PHY-1125915. JPL was supported by a grant from the French Space Agency CNES.
\end{acknowledgements}

\end{document}